\documentclass[]{mn2e}
\usepackage{graphicx}

\title[The dust mass in $z > 6$ normal star forming galaxies]{The dust mass in $z > 6$ normal star forming galaxies}
\author[Mancini et al.]{Mattia Mancini$^{1,2}$\thanks{E-mail:
mattia.mancini@oa-roma.inaf.it}, Raffaella Schneider$^{1}$, Luca Graziani$^{1}$, Rosa Valiante$^{1}$,
\newauthor 
Pratika Dayal$^{3}$, Umberto Maio$^{4,5}$, Benedetta Ciardi$^{6}$ and Leslie K. Hunt$^7$\\
$^{1}$INAF/Osservatorio Astronomico di Roma, Via di Frascati 33, 00040 Monte Porzio Catone, Italy\\
$^{2}$Dipartimento di Fisica, ``Sapienza'' Universit{\'a} di Roma, Piazzale Aldo Moro 5, 00185, Roma, Italy\\
$^{3}$  Institute for computational cosmology, University of Durham, South Road, Durham, DH1 3LE, UK   \\
$^{4}$ INAF - Osservatorio Astronomico di Trieste, via G. B. Tiepolo 11, 34131 Trieste, Italy \\
$^{5}$Leibniz Institute for Astrophysics, an der Sternwarte 16, 14482 Potsdam, Germany\\
$^{6}$Max Planck Institut f¨ur Astrophysik, Karl-Schwarzschild-Strasse 1, 85741 Garching, Germany   \\
$^7$ INAF/Osservatorio Astrofisico di Arcetri, Largo Enrico Fermi 5, 50125 Firenze, Italy }
\begin{document}

\date{29 April 2015}

\pagerange{\pageref{firstpage}--\pageref{lastpage}} \pubyear{2015}

\maketitle
\label{firstpage}

\begin{abstract}
We interpret recent ALMA observations of $z > 6$ normal star forming galaxies by means
of a semi-numerical method, which couples the output of a cosmological hydro-dynamical simulation
with a chemical evolution model which accounts for the contribution to dust enrichment from supernovae,
asymptotic giant branch stars and grain growth in the interstellar medium.
We find that while stellar sources dominate the dust mass of small galaxies, the higher level of metal
enrichment experienced by galaxies with $\rm M_{star} > 10^9 \, M_\odot$ allows efficient grain growth,
which provides the dominant contribution to the dust mass. Even assuming
maximally efficient supernova dust production, the observed dust mass of the $z = 7.5$ galaxy A1689-zD1 requires very efficient grain growth.
This, in turn, implies that in this galaxy the average density of the cold and dense gas,
where grain growth occurs, is comparable to that inferred from observations of QSO host
galaxies at similar redshifts. Although plausible, the upper limits on the dust continuum
emission of galaxies at $6.5 < z < 7.5$ show that these conditions must not apply to the bulk of
the high redshift galaxy population.
\end{abstract}


\section{Introduction}

Observations at millimeter (mm) and sub-millimeter (sub-mm) wavelengths have
provided convincing evidence of rapid dust enrichment at high redshift (for a recent review 
see Carilli \& Walter 2013 and references therein).
Dust masses  as large as  $\rm \sim 10^8 M_{\odot}$ have been inferred for $5 \le z \le 7$ QSO host galaxies, 
requiring a very efficient dust formation channel that must operate in less than 1 Gyr of cosmic evolution.
Theoretical studies have shown that although stellar sources of dust, including both
Supernovae (SN) and Asymptotic Giant Branch (AGB) stars, can be fast enough to significantly contribute to high redshift dust enrichment 
(Valiante et al. 2009), the observed dust masses require efficient grain growth in the dense phase
of the interstellar medium (ISM, Michalowski et al. 2010; Valiante et al. 2011; 2014).
Until very recently, sub-mm/mm continuum observations at $z \sim 6-7$ have been limited to ``extreme'' galaxies,
QSO hosts or strong starbursts, which are characterized by star formation rates ranging from
several hundreds to thousands of solar masses per year. However, at these high redshifts we expect the bulk of the galaxy
population to be characterized by ``normal'' galaxies, which form stars at rates from a few to a few tens of solar masses per year. 
Deep ALMA and Plateau de Bure Interferometer (PdBI) follow up observations have mostly provided upper
limits on the dust mass of such galaxies (Kanekar et al. 2013; Ouchi et al. 2013; Ota et al. 2014; Schaerer et al. 2015;  Maiolino et al. 2015),
with the exception of the gravitationally lensed $z = 7.5$ Lyman Break Galaxy (LBG) A1689-zD1, whose dust continuum emission has been
recently detected with ALMA (Watson et al. 2015). The estimated dust mass is  $\rm 4^{+4}_{-2} \times 10^7 M_\odot$ ,
showing that dusty galaxies have already formed at $z > 7$.

In this paper, we investigate early dust enrichment in ``normal'' star forming galaxies at $z \ge 6$. We use the output of a
cosmological hydro-dynamical simulation which allows the prediction of the gas, stellar and metal content of galaxies along their hierarchical
assembly (Maio et al. 2010). We then post-process the simulation output with a semi-analytical chemical evolution model 
with dust (Valiante et al. 2009; 2011; 2014; de Bennassuti et al. 2014) to estimate their dust masses. A similar
semi-numerical approach has been followed by Dayal, Hirashita \& Ferrara (2010), who considered the contribution to
dust enrichment of SN and estimate the Far InfraRed  (FIR)  detectability
of high-redshift galaxies classified as Lyman-$\alpha$ emitters (LAE). More recently, Hirashita et al. (2014) 
have been able to put constraints on the dust production rate by SN using the upper limit on the dust
continuum emission of Himiko, one of the best-studied LAE at $z \sim 6.6$ (Ouchi et al. 2013). Here we attempt to improve on 
these previous studies by considering dust enrichment from SN, AGB stars
and grain growth in the ISM with the aim of assessing their relative contribution to the total mass of dust in the ISM of $z \ge 6$
galaxies. We compare the model predictions with currently available upper limits on the dust mass at $6 \le z \le 7.5$ and
discuss the implications posed on the models by the newly discovered dusty galaxy at $z = 7.5$ (Watson et al. 2015).

The paper is organized as follows: in section \ref{sec:observed} we introduce the sample of observed galaxies that we 
have collected from literature papers; in section \ref{sec:model} we give a brief presentation of the semi-numerical 
model and in section \ref{sec:results} we discuss the main results. Finally, in section \ref{sec:conclusions} we 
draw our main conclusions.

\begin{table}

  \caption{Physical properties of the galaxy sample collected from the literature. Lensed objects have been corrected using the appropriate magnification factor ($\rm \mu = 4.5, 9\, and\, 9.3$ for  HCM6A, A1703-zD1 and A1689-zD1, respectively).
  All the quantities are computed assuming conversion factors, scaling relations and dust properties
  presented in the text.
  References: $a$: Schaerer et al. (2015);  $b$: Kanekar et al. (2013); $c$: Ota et al. (2014); $d$: Ouchi et al. (2013); $e$: Maiolino et al. (2015); $f$: Watson et al. (2015).}
  \label{tab:sample}
  \begin{tabular} {|l|c|c|c|c} \hline

  Name    	             &    z        & $\rm M_{UV}$       &  $\rm Log \, M_{star}$   &  $\rm Log \, M_{\rm dust}$     \\ 
                	             &              &     mag               &  $\rm [M_\odot]$  &  $\rm [M_\odot]$               \\ \hline
                	             
 A1703-zD1$^a$    & 6.800     &    -20.3               &     $9.2 \pm 0.3 $     &    $< 7.36 $                  \\  
z8-GND-5296$^a$ & 7.508    &    -21.4	             &     $9.7 \pm 0.3$	  &    $< 8.28  $                 \\
HCM6A$^b$           & 6.560    &   -20.8            &     $9.5 \pm 0.3$      &    $< 7.61  $                    \\
IOK-1$^c$              &  6.960   &   -21.3             &     $9.7  \pm 0.3$     &    $ < 7.43  $              \\
Himiko$^d$             & 6.595   &   -21.7                &     $9.9 \pm 0.3 $      &    $ <7.30  $               \\\hline
BDF-3299$^e$       & 7.109    &   -20.44            &     $9.3\pm 0.3 $      &    $ <7.02  $              \\
BDF-512$^e$         &7.008     &   -20.49              &     $9.3 \pm 0.3$       &	  $<7.36  $             \\
SDF-46975$^e$     &6.844     &   -21.49          &     $9.8 \pm 0.3$       &	  $ <7.38 $               \\ \hline
A1689-zD1$^f$	    & 7.500     &    -19.7             &     $ 9.0 \pm 0.3$      &      $ 7.51 \pm 0.2$                \\ \hline
\end{tabular}
\end{table}

\section{The observed sample}
\label{sec:observed}

In Table \ref{tab:sample} we have summarized the properties of the sample collected from the literature. For 8 out of the 9 galaxies, upper limits on the dust continuum emission have been obtained by means of deep ALMA and PdBI observations (Kanekaer et al. 2013; Ouchi et al. 2013; Ota et al. 2014; Schaerer et al. 2015; Maiolino et al. 2015).
The  $z = 7.5$ LBG  A1689-zD1 is currently the most distant UV-selected galaxy for which
a dust continuum detection has been obtained with ALMA (Watson et al. 2015). 
For all these galaxies, the star formation rates estimated  from the UV luminosity\footnote{ We have adopted a conversion factor of
$\rm  {\cal K}_{UV} = \rm 1.15 \times 10^{-28} \, M_\odot \, yr^{-1}/erg \, s^{-1} Hz^{-1}$, as appropriate for a stellar population with metallicity $\rm 0.3 \, Z_\odot < Z < \, Z_\odot$ formed at a constant SFR
with a Salpeter IMF in the range $\rm [0.1 - 100] \, M_\odot$ at age $\rm \sim 300 \, Myr$ (Madau \& Dickinson 2014).}
range between $4.1$ and $\rm 22.4\, M_\odot/yr$.  
Following {\it Schaerer et al. (2015)},
we derive the stellar masses using the mean relation between the UV magnitude and the stellar mass obtained by the same authors from detailed
fits of the spectral energy distribution of a sample of $z \sim 6.7$ LBGs, including nebular emission and dust attenuation:
\begin{equation}
\rm Log (M_{star}/M_\odot) = -0.45 \times (M_{UV} + 20) + 9.11.
\end{equation}
\noindent
Dayal et al. (2014) obtain a similar $\rm M_{star} - M_{UV}$ relation on the basis of a theoretical model,
which also accounts for the rapid decline of the UV luminosity with stellar ages.

Assuming that the dust is optically thin in the rest-frame FIR, 
the upper limits on the dust mass have been obtained taking into account the
effect of the Cosmic Microwave Background (CMB) on the intrinsic dust emission (da Cunha et al. 2013).
%
In Table \ref{tab:sample} we report dust masses adopting a dust temperature\footnote{$\rm T_{dust,0}$ is the dust temperature heated by a stellar radiation field at $z=0$ (da Cunha et al. 2013).} 
$\rm T_{dust,0} = 35~K$ and a dust emissivity $\rm k_{\nu_{res}} = \rm  k_0 (\lambda_0/\lambda_{res})^\beta$
with $\rm k_0 =   0.77\, cm^2/gr$, $\lambda_0 = 850 \,\mu$m and $\beta = 1.5$ (Ota et al. 2014). 
The dust mass can increase/decrease by $\rm \sim 0.4 \,dex$ for variations in the dust temperature in the range  $\rm 25 K \le T_{dust,0} \le 45 K$. In addition,
the poorly constrained dust properties make the value of the emissivity coefficient to be adopted very uncertain (see Table 1 in Hirashita et al. 2014). At fixed
dust temperature, we find that variations in $\rm k_\nu$ among values adopted in the literature (Weingartner \& Draine 2001) or
applied to submm observations of high-$z$ galaxies (Michalowski et al. 2010; Valiante et al. 2014; Watson et al. 2015) introduce an additional  $\rm \sim 0.3 \,dex$ uncertainty in the estimated dust mass.
\begin{figure}
\vspace{\baselineskip}
\includegraphics[width=8cm]{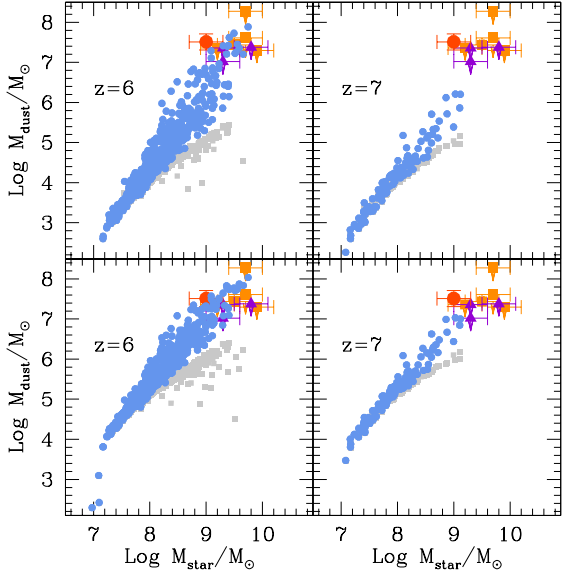}
\caption{Predicted dust masses of the simulated galaxies as a function of the stellar mass. For each galaxy, the dust mass without (with) grain growth
is shown by a square grey (circle blue) point (see text). The adopted grain growth timescale is
$\rm \tau_{acc,0} = 2 \, Myr$. In the lower panels, the reverse shock destruction of SN dust is neglected.
For the sake of comparison, we have reported the same data points shown in Table 1 in the two panels: Schaerer et al. (2015,
squares), Maiolino et al. (2015, triangles) and Watson et al. (2015, circle point).
}
 \label{fig:dustmass}
 \end{figure}

\section{The model}
\label{sec:model}

We briefly describe the simulation used in this work and interested readers are referred to Maio et al. (2010), 
Campisi et al. (2011), Salvaterra et al. (2013) and Dayal et al. (2013) for more details. 
The smoothed particle hydrodynamic (SPH) simulation is carried out using the TreePM-SPH code 
{\small {GADGET-2}} (Springel 2003) and has a periodic comoving box size of $\rm 30 \,h^{-1} cMpc$. 
It initially contains $2\times 320^3$ particles, with the masses of gas and DM particles being 
$\rm 9\times 10^6 h^{-1} M_\odot$ and $\rm 6 \times 10^7 h^{-1} M_\odot$, respectively\footnote{The cosmological model corresponds
to the $\Lambda$CDM Universe with DM, dark energy and baryonic density parameter values of
($\Omega_{\rm m },\Omega_{\Lambda}, \Omega_{\rm b}) = (0.3,0.7,0.04)$, a Hubble constant $H_0 = 100h= 70 {\rm km\, s^{-1} Mpc^{-1}}$,
a primordial spectral index $n_s=1$ and a spectral normalisation $\sigma_8=0.9$. }.
The code includes the cosmological evolution of both atomic and ionized Hydrogen, Helium and Deuterium (e.g. Yoshida et al. 2003;  Maio et al. 2009), 
PopIII and PopII/I star formation according to the corresponding initial mass function (IMF; Tornatore et al. 2007), 
gas cooling from resonant and fine-structure lines (Maio et al. 2007) and
feedback effects assuming a multi-phase ISM (Springel \& Hernquist 2003). 
The chemical model follows the detailed evolution of each SPH particle and the abundances of different 
elemental species are consistently derived at each timestep using the lifetime function 
(Padovani \& Matteucci 1993) and metallicity-dependant stellar yields from core-collapse SN, AGB stars, SNIa and pair instability SN.

Galaxies are recognized as gravitationally-bound groups of at least 32 total (DM+gas+star) particles
by running a friends-of-friends (FOF) algorithm. Of the galaxies identified
in the simulation, in our calculations we only use `well-resolved' galaxies that contain at
least 10 star particles, corresponding to more than 145 total particles at any redshift. 
As a result, the number of galaxies is 1000 at $z \sim 7$ and 1863 at $z \sim 6$.

For each galaxy, we compute the intrinsic spectral energy distribution (SED) summing the contribution
of all its stellar populations by means of the Starburst99 code (Leitherer et al. 1999; Vazquez \& Leitherer 2005).
In doing so, we consider the ages and metallicities of each stellar population as predicted by the numerical simulation.
Then we convert the luminosity at $\rm \lambda_{res} = 1500 \,\AA$ into absolute UV magnitudes, $\rm M_{UV}$. 
Given the UV magnitudes of the galaxies in the observational sample, in what follows we will focus only on the brightest
galaxies in the simulation box, selecting the ones with $\rm M_{UV} \le -18$ (102 objects at $z \sim 7$ and 225 at $z\sim 6$).
The metallicities and stellar ages of these galaxies increase with the stellar mass, with  $\rm 5\times 10^{-3} \, Z_\odot \le Z \le 0.1 \, Z_{\odot}$
and $\rm 40 \, Myr \le t_{age} \le 200 \, Myr$ for $\rm 7.5 \le Log\, M_{star}/M_\odot \leq 9.5$. In addition, their star formation rates are $\rm \le 30 \, M_\odot/yr$, 
hence they can be considered as ``normal'' galaxies. \\

\begin{figure}
\vspace{\baselineskip}
\includegraphics[width=8cm]{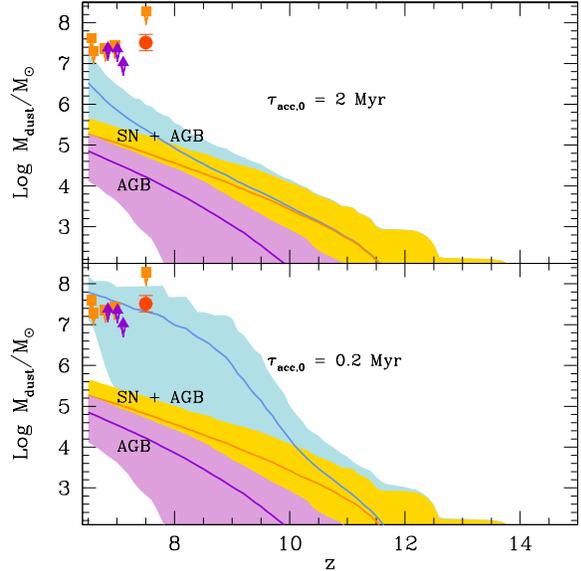}
\caption{Redshift evolution of the dust mass for the simulated galaxies with stellar masses in the range $\rm Log \, M_{star}/M_{\odot} \ge 9$. 
Each line represents the average contribution of all the galaxies with the shaded area indicating the dispersion among different evolutionary
histories. The lower, intermediate and upper lines show the contribution to the total mass of dust of AGB stars, stellar sources and grain growth
with an accretion timescale of $\rm \tau_{acc,0} = 2 \, Myr$ (upper panel) and 0.2 Myr (lower panel).}
 \label{fig:dustevo}
 \end{figure}
\noindent
The dust content of a galaxy depends on its star formation and metal enrichment histories. 
Since dust evolution is not computed within the 
simulation, our approach was to reconstruct the star formation history of each simulated galaxy at $z = 6$ and $z=7$ along its hierarchical assembly
and then use this as an input to the chemical evolution model with dust. This model has been first developed by Valiante et al. (2009), then applied
to semi-analytical merger tree models of high-redshift QSOs by Valiante et al. (2011, 2012, 2014) and recently improved by de Bennassuti et al. (2014)
to explain the metal and dust content of the Milky Way and its progenitors. We refer the interested readers to these 
papers for more details. 

The equation describing the dust mass evolution is, 

\begin{eqnarray}
 \rm \dot{M}_{d}(t)  & = &  \rm - Z_d(t) \rmn{SFR}(t) + \dot{Y}_{d}(t)  \nonumber \\
 & & \rm  - (1- X_c) \frac{M_{d}(t)}{\tau_{d}}+ X_c \frac{M_{d}(t)}{\tau_{acc}} 
\label{eq:dustevo}   
\end{eqnarray}
\noindent
where $\rm M_d$ is the dust mass, $\rm Z_{d}(t)=M_{d}(t)/M_{\rm ISM}(t)$ is the total dust abundance in the ISM, $\rm X_c$ is the
cold gas mass fraction and $\rm \tau_{d}$ and $\rm \tau_{acc}$ are the timescales for grain destruction and grain growth and will be
discussed below.

The time-dependent term, $\rm \dot{Y}_d(t)$, is the dust production rate and it
depends on the stellar IMF and on the adopted model for the stellar yields. We compute it as follows:
\begin{equation}
\rm \dot{Y}_{d}(t) = \int_{m(t)}^{m_{up}} m_d(m,Z)\ \Phi(m)\ SFR(t - \tau_m)\ dm,
\label{eq:dustyields}
\end{equation}
\noindent
where $\rm m_d$ is the dust mass yield, 
which depends on the stellar mass and 
metallicity, $\rm \Phi(m)$ is the stellar IMF, and the lower limit of integration, $\rm m(t)$, is the mass of a star with a lifetime $\rm \tau_m = t$. 
To be consistent with the assumptions made in the simulation,
we adopt a Salpeter law in the mass range 
$\rm 0.1\, M_\odot \le m \le 100 \, M_\odot$\footnote{Since Pop III stars in the simulated 
galaxies are always largely subdominant, we neglect their contribution to chemical enrichment (Dayal et al. 2010).}.
For stars with $\rm m  < 8\, M_\odot$, 
we adopt metal yields from
Van den Hoek \& Groenewegen (1997) 
and dust yields  by 
Zhukovska et al. (2008). 
For massive stars $\rm (12\, M_\odot < m < 40\, M_\odot)$ 
metal and dust yields have been taken from 
Woosley \& Weaver (1995) 
and from Bianchi \& Schneider (2007), 
including the effect of the reverse shock on dust survival
in SN ejecta. 
For stars in the intermediate mass range $\rm 8\, M_\odot < m  < 12\, M_\odot$, 
we interpolate between the AGB yields 
for the largest-mass progenitor and SN yields from the lowest-mass progenitor.
Above $\rm 40 \, M_\odot$, stars are assumed to collapse to black hole without contributing to the enrichment of the ISM. 

Finally, the last two terms in the right-hand side of eq.~\ref{eq:dustevo} represent the effects of dust destruction by interstellar
shock waves and grain growth in the dense phase of the ISM. Here we 
assume that at any given time, a fixed fraction $\rm X_c =0.5$ of ISM is in the cold phase\footnote{
The value of $\rm X_c$, averaged over simulated galaxies with different masses, ranges between 0.38 and 0.56 (Vallini, Dayal \& Ferrara 2012).}.
The timescale for grain destruction, $\rm \tau_{d}$, and grain growth, $\rm \tau_{acc}$, are computed as in 
de Bennassuti et al. (2014). 
The latter timescale
depends on the grain size distribution, and on the gas phase metallicity $\rm Z$, temperature $\rm T_{mol}$, and density $\rm n_{mol}$, of the cold molecular phase where grain growth is most efficient,
\begin{eqnarray}
\tau_{\rm acc} & = & \rm 20 \, Myr \times \Big( \frac{n_{\rm mol}}{100\, cm^{-3}} \Big)^{-1}\, \Big( \frac{T_{\rm mol}}{50\, K} \Big)^{-1/2}\, \Big( \frac{Z}{Z_\odot} \Big)^{-1} \nonumber \\
& = &\rm  \tau_{acc,0}  \rm \, \Big( \frac{Z}{Z_\odot} \Big)^{-1}
\label{eq:acc}
\end{eqnarray}
\noindent
where we have assumed that grains which experience grain growth have a typical  size of $\sim 0.1\,\mu$m (Hirashita et al. 2014). For gas at solar metallicity,
with $\rm n_{mol} = \rm 10^3 cm^{-3}$ and $\rm T_{\rm mol} = \rm 50 \, K$, the accretion timescale is $\rm  \tau_{acc,0} =  2 \,Myr$ (Asano et al. 2013).  
Smaller grain sizes increase the grain surface area per unit dust mass,
thus shortening the accretion timescale (Kuo \& Hirashita 2012).
Since the simulation has not
the resolution to trace the temperature and density of the ISM, the values of $\rm n_{\rm mol}$ and $\rm T_{\rm mol}$ which best describe the galaxies under investigation
are unconstrained. For this reason, in what follows we explore the impact of different values of $\rm \tau_{acc,0}$ on the resulting dust masses.

\section{Results}
\label{sec:results}

Figure \ref{fig:dustmass} shows a comparison between the predicted dust masses of the simulated galaxies and the observations.
We show the dust mass as a function of the stellar mass for all the simulated galaxies with intrinsic $\rm M_{UV} \le -18$ at $z=6.33$ and $z=7.14$. 
For each galaxy, the light grey point represents the mass of dust that is produced by stellar sources (AGB stars and SNe) after
grain destruction and astration (eq.~\ref{eq:dustevo} without grain growth), and the blue point is
the total dust mass, including the effect of grain growth in the ISM. Here we have assumed a grain growth timescale $\rm \tau_{acc,0} = 2\, Myr$.
Due to the lower metallicities, 
grain growth is not efficient at stellar masses $\rm M_{\rm star} < 10^8 M_{\odot}$. In these small galaxies, we predict that all the existing
dust mass is entirely contributed by stellar sources. Galaxies with larger stellar masses have experienced a larger degree of chemical enrichment
and the gas metallicities are high enough to activate efficient grain growth, although with a large scatter.   
Yet, due to the time required to enrich the ISM with metals, the bulk of the dust mass in the most massive galaxies is grown between $z \sim 7$ and $z \sim 6$. In fact, while at $z \sim 7$
all the simulated galaxies have $\rm M_{dust} < 2 \times 10^6 \, M_{\odot}$, at $z \sim 6$ there are several galaxies
with dust masses $\rm 10^7 M_{\odot} \le M_{dust} \le 10^8\, M_{\odot}$. Note that the observational data points, that we have reported in both
panels for the sake of comparison, refer to galaxies which span a redshift range $6.5 \le z \le 7.5$ (see Table 1). 

These results depend on 
the adopted stellar dust yield and grain growth timescale. 
In the lower panel of Figure \ref{fig:dustmass} we show
the effect of reducing the efficiency of grain destruction by the SN reverse shock: if all the grains survive the passage of the reverse shock,
the effective SN dust yields is $\sim 20$ times larger and the resulting dust masses contributed by stellar sources increase by a comparable factor.
These SN yields bracket the observations of dust masses in SN and SN remnants obtained with the {\it Herschel} satellite
(see Fig.~6 in Schneider et al. 2014). The figure shows that although the total dust mass is larger, even with maximally efficient SN dust enrichment 
the predicted dust masses at $z \sim 7$ are $\rm M_{\rm dust} < 10^7 M_{\odot}$, too small to account for the observed dust mass in
A1689-zD1. We conclude that in order to account for the existing dust mass in this galaxy, a shorter grain growth timescale is required.

It is interesting to analyse the  relative contribution of AGB stars, SN and grain growth as a function of redshift. Figure \ref{fig:dustevo} shows the
redshift evolution of the dust mass for galaxies with 
$\rm M_{star} > 10^{9} \,M_{\odot}$. We separate the contribution from AGB stars, stellar sources, and the total mass of dust, including 
grain growth with an accretion timescale $\rm \tau_{acc,0} = 2 \, Myr$ (upper panel) and 0.2 Myr (lower panel). 
As expected from Figure~\ref{fig:dustmass}, grain growth provides the dominant contribution, exceeding the dust produced by stellar sources already at $z < 10 - 12$.
Among the stellar sources, SN appear always dominant, but the average contribution of AGB stars can be as large as
 $\sim 40\%$. This confirms that the contribution of AGB stars 
to high redshift dust formation can not be neglected, especially for the galaxies currently targeted by observational searches (Valiante et al. 2009). 
Finally, in the lower panel we show that the dust mass detected in A1689-zD1 requires very efficient grain growth, with a timescale 
$\rm \tau_{acc,0} = 0.2 \, Myr$, one order of magnitude shorter than required to reproduce the observed dust-to-gas ratio in the Milky Way
and in most local dwarf galaxies (de Bennassuti et al. 2014).
This, in turn, implies that the  cold atomic and molecular phases of the ISM, where grain growth is more efficient, must have an average density
of $\rm \sim 10^4 cm^{-3}$ (see eq.~\ref{eq:acc}). 
Such a value is comparable to the molecular gas density inferred from CO
excitation analyses of starburst galaxies at comparable (although slightly smaller)
redshifts (see Carilli \& Walter 2013). Morever, the molecular phase in these high-redshift star-forming galaxies may
also be warmer, leading to somewhat lower $\tau_{\rm acc,0}$.
Although plausible, our study suggests that 
these conditions must be exceptional, as if they were to apply to all galaxies a $z > 6.5$,  current upper limits on the dust continuum emission for normal star forming galaxies
at $6.5 < z < 7.5$ would be exceeded.    


\section{Conclusions}
\label{sec:conclusions}
In this work we have investigated dust enrichment of the ISM of ``normal'' galaxies, which form stars at rates between a few to a few tens of solar masses per year, at $z \ge 6$. We find that dusty galaxies are already formed at these high redshifts,
due to rapid enrichment by SN and AGB stars. However, the dominant contribution to the dust mass in the most massive galaxies, with $\rm  M_{star} \ge 10^9 M_\odot$,
which are the galaxies detected in the UV with present facilities, comes from grain growth. The efficiency of this process depends on the metallicity, density and temperature
of the cold and dense phase of the ISM.  While the upper limits on the dust continuum emission can be matched with an accretion timescale of
$\rm \tau_{acc,0} \sim 2 \, Myr$,  comparable to the value adopted by de Bennassuti
et al. (2014) to reproduce the dust-to-gas ratio of the Milky Way and local dwarfs, the observed dust mass of A1689-zD1 at $z = 7.5$ can only be  reproduced if
more extreme conditions are adopted, with $\rm \tau_{acc,0} = 0.2 \, Myr$, similar to the value assumed for QSO host galaxies (Valiante et al. 2014). 
It is interesting to note that physical conditions in the ISM can lead to different values of $\rm \tau_{acc, 0}$ even for
galaxies with comparable metallicity: two of the most metal-poor local dwarfs, SBS 0335-052 and IZw18, have a metallicity of $\rm Z \sim 3 \% \, Z_\odot$
but widely different dust masses (Hunt et al. 2014). The proposed interpretation is that the greater dust mass in SBS 0335-052 is due to the more efficient grain
growth allowed by the high density ISM, observationally inferred to be almost 20 times higher than in IZw18 (Schneider, Hunt \& Valiante 2015).
We conclude that, when the ISM is characterized by large enough densities, dust masses in high redshift galaxies may be more common than hitherto believed
and that the level of dust enrichment may not always be satisfactorily assessed from the gas metallicity.

\section*{Acknowledgments}
We thank the Referee, Hiroyuki Hirashita, for his insightful comments, Laura Pentericci and Daniel Schaerer for their kind clarifications.
The research leading to these results has received funding from the European Research Council under the European 
Union 
(FP/2007-2013) / ERC Grant Agreement n. 306476.
PD acknowledges the support of the Addison Wheeler Fellowship awarded by the Institute of Advanced Study at Durham University. UM has been funded through a Marie Curie fellowship by the European Union 
(FP/2007-2013), grant agreement n. 267251.

\label{lastpage}

\end{document}